# CMOS-compatible multi-band plasmonic TE-pass polarizer

**NICOLÁS ABADÍA,**[1,2,3,4,5,6,*] **MD. GHULAM SABER,**[1] **FRANK BELLO,**[2,3,4] **ALIREZA SAMANI,**[1] **ESLAM EL-FIKY,**[1] **YUN WANG,**[1] **JOHN F. DONEGAN,**[2,3,4] **AND DAVID V. PLANT**[1]

[1]*Photonic Systems Group, Department of Electrical and Computer Engineering, McGill University, 3480 University Street, Montréal H3A 0E9, Canada*
[2]*Photonics Group, School of Physics, Trinity College, Dublin 2, Ireland*
[3]*Centre for Research on Adaptive Nanostructures and Nanodevices (CRANN), Trinity College Dublin, Dublin 2, Ireland*
[4]*AMBER Research Centre, Trinity College Dublin, Dublin 2, Ireland*
[5]*School of Physics and Astronomy, Cardiff University, Queen's Buildings, The Parade, Cardiff CF24 3AA, UK*
[6]*Institute for Compound Semiconductors, Cardiff University, Queen's Buildings, The Parade, Cardiff CF24 3AA, UK*
[*]*abadian@cardiff.ac.uk*

**Abstract:** A CMOS-compatible plasmonic TE-pass polarizer capable of working in the O, E, S, C, L, and U bands is numerically analyzed. The device is based on an integrated hybrid plasmonic waveguide (HPW) with a segmented metal design. The segmented metal will avoid the propagation of the TM mode, confined in the slot of the HPW, while the TE fundamental mode will pass. The TE mode is not affected by the metal segmentation since it is confined in the core of the HPW. The concept of the segmented metal can be exploited in a plasmonic circuit with HPWs as the connecting waveguides between parts of the circuit and in a silicon photonics circuit with strip or slab waveguides connecting the different parts of the circuit. Using 3D FDTD simulations, it is shown that for a length of 5.5 μm the polarization extinction ratios are better than 20 dB and the insertion losses are less than 1.7 dB over all the optical communication bands.

## References and links

1. N. Abadía, X. Dai, Q. Lu, W.-H. Guo, D. Patel, D. V. Plant, and J. F. Donegan, "Highly fabrication tolerant InP based polarization beam splitter based on p-i-n structure," Opt. Express **25**(9), 10070–10077 (2017).
2. N. Abadia, Xiangyang Dai, Q. Lu, W.-H. Guo, E. El-Fiky, D. V. Plant, and J. F. Donegan, "Novel polarization beam splitter based on p-i-n structure for an indium phosphide platform," in 2017 19th International Conference on Transparent Optical Networks (ICTON, 2017), paper Tu.A5.1.
3. Y. Huang, S. Zhu, H. Zhang, T.-Y. Liow, and G.-Q. Lo, "CMOS compatible horizontal nanoplasmonic slot waveguides TE-pass polarizer on silicon-on-insulator platform," Opt. Express **21**(10), 12790–12796 (2013).
4. K. Takiguchi, K. Okamoto, S. Suzuki, and Y. Ohmori, "Planar lightwave circuit optical dispersion equalizer," IEEE Photonics Technol. Lett. **6**(1), 86–88 (1994).
5. C. R. Doerr, K. W. Chang, L. W. Stulz, R. Pafchek, Q. Guo, L. Buhl, L. Gomez, M. Cappuzzo, and G. Bogert, "Arrayed waveguide dynamic gain equalization filter with reduced insertion loss and increased dynamic range," IEEE Photonics Technol. Lett. **13**(4), 329–331 (2001).
6. K. Takiguchi, H. Takahashi, and T. Shibata, "Tunable chromatic dispersion and dispersion slope compensator using a planar lightwave circuit lattice-form filter," Opt. Lett. **33**(11), 1243–1245 (2008).
7. K. Takiguchi, K. Jinguji, K. Okamoto, and Y. Ohmori, "Variable Group-Delay Dispersion equalizer using lattice-form programmable optical filter on planar lightwave circuit," IEEE J. Sel. Top. Quantum Electron. **2**(2), 270–276 (1996).
8. J. S. Orcutt, B. Moss, C. Sun, J. Leu, M. Georgas, J. Shainline, E. Zgraggen, H. Li, J. Sun, M. Weaver, S. Urošević, M. Popović, R. J. Ram, and V. Stojanović, "Open foundry platform for high-performance electronic-photonic integration," Opt. Express **20**(11), 12222–12232 (2012).
9. D. Thomson, A. Zilkie, J. E. Bowers, T. Komljenovic, G. T. Reed, L. Vivien, D. Marris-Morini, E. Cassan, L.

Virot, J.-M. Fédéli, J.-M. Hartmann, J. H. Schmid, D.-X. Xu, F. Boeuf, P. O'Brien, G. Z. Mashanovich, and M. Nedeljkovic, "Roadmap on silicon photonics," J. Opt. **18**(7), 073003 (2016).
10. N. Abadía, T. Bernadin, P. Chaisakul, S. Olivier, D. Marris-Morini, R. Espiau de Lamaëstre, J. C. Weeber, and L. Vivien, "Low-Power consumption Franz-Keldysh effect plasmonic modulator," Opt. Express **22**(9), 11236–11243 (2014).
11. N. Abadía, S. Olivier, D. Marris-Morini, L. Vivien, T. Bernadin, and J. C. Weeber, "A CMOS-compatible Franz-Keldysh effect plasmonic modulator," in 2014 11th International Conference on Group IV Photonics (GFP, 2014), pp. 63–64.
12. M. A. Foster, A. C. Turner, R. Salem, M. Lipson, and A. L. Gaeta, "Broad-band continuous-wave parametric wavelength conversion in silicon nanowaveguides," Opt. Express **15**(20), 12949–12958 (2007).
13. Y. Xiong, D. X. Xu, J. H. Schmid, P. Cheben, and W. N. Ye, "High extinction ratio and broadband silicon TE-pass polarizer using subwavelength grating index engineering," IEEE Photonics J. **7**(5), 7802107 (2015).
14. G. V. Naik, V. M. Shalaev, and A. Boltasseva, "Alternative plasmonic materials: Beyond gold and silver," Adv. Mater. **25**(24), 3264–3294 (2013).
15. M. Alam, J. S. Aitchison, and M. Mojahedi, "Compact hybrid TM-pass polarizer for silicon-on-insulator platform," Appl. Opt. **50**(15), 2294–2298 (2011).
16. Y. Xu and J. Xiao, "A compact TE-pass polarizer for silicon-based slot waveguides," IEEE Photonics Technol. Lett. **27**(19), 2071–2074 (2015).
17. Z. Ying, G. Wang, X. Zhang, Y. Huang, H.-P. Ho, and Y. Zhang, "Ultracompact TE-pass polarizer based on a hybrid plasmonic waveguide," IEEE Photonics Technol. Lett. **27**(2), 201–204 (2015).
18. X. Guan, P. Chen, S. Chen, P. Xu, Y. Shi, and D. Dai, "Low-loss ultracompact transverse-magnetic-pass polarizer with a silicon subwavelength grating waveguide," Opt. Lett. **39**(15), 4514–4517 (2014).
19. X. Sun, M. Z. Alam, S. J. Wagner, J. S. Aitchison, and M. Mojahedi, "Experimental demonstration of a hybrid plasmonic transverse electric pass polarizer for a silicon-on-insulator platform," Opt. Lett. **37**(23), 4814–4816 (2012).
20. M. Z. Alam, J. S. Aitchison, and M. Mojahedi, "Compact and silicon-on-insulator-compatible hybrid plasmonic TE-pass polarizer," Opt. Lett. **37**(1), 55–57 (2012).
21. X. Guan, P. Xu, Y. Shi, and D. Dai, "Ultra-compact broadband TM-pass polarizer using a silicon hybrid plasmonic waveguide grating," in Asia Communications and Photonics Conference (2013), pp. 11–13.
22. T. Davis, D. Gómez, and A. Roberts, "Plasmonic circuits for manipulating optical information," Nanophotonics **6**, 543–559 (2017).
23. T. K. Ng, M. Z. M. Khan, A. Al-Jabr, and B. S. Ooi, "Analysis of CMOS compatible Cu-based TM-pass optical polarizer," IEEE Photonics Technol. Lett. **24**(9), 724–726 (2012).
24. S. I. Azzam and S. S. A. Obayya, "Titanium nitride-based CMOS-compatible TE-pass and TM-pass plasmonic polarizers," IEEE Photonics Technol. Lett. **28**(3), 367–370 (2016).
25. Y. Xu and J. Xiao, "Design and numerical study of a compact, broadband and low-loss TE-pass polarizer using transparent conducting oxides," Opt. Express **24**(14), 15373–15382 (2016).
26. T. Huang, "TE-pass polarizer based on epsilon-near-zero material embedded in a slot waveguide," IEEE Photonics Technol. Lett. **28**(20), 2145–2148 (2016).
27. R. F. Oulton, V. J. Sorger, D. A. Genov, D. F. P. Pile, and X. Zhang, "A hybrid plasmonic waveguide for subwavelength confinement and long-range propagation," Nat. Photonics **2**(8), 496–500 (2008).
28. M. Fukuda, H. Sakai, T. Mano, Y. Kimura, M. Ota, M. Fukuhara, T. Aihara, Y. Ishii, and T. Ishiyama, "Plasmonic and electronic device integrated circuits and their characteristics," in 2015 45th European Solid State Device Research Conference (ESSDERC, 2015), pp. 105–108.
29. A. Emboras, A. Najar, S. Nambiar, P. Grosse, E. Augendre, C. Leroux, B. de Salvo, and R. E. de Lamaestre, "MNOS stack for reliable, low optical loss, Cu based CMOS plasmonic devices," Opt. Express **20**(13), 13612–13621 (2012).
30. I. Lumerical Solutions, "Lumerical Solutions, Inc. http://www.lumerical.com," http://www.lumerical.com.
31. S. Maier, Plasmonics: Fundamentals and applications (Springer, 2007).
32. M. G. Saber, N. Abadía, and D. V. Plant, "CMOS compatible all-silicon TM pass polarizer based on highly doped silicon waveguide," Opt. Express **26**(16), 20878–20887 (2018).

## 1. Introduction

For many applications of silicon photonic integrated circuits, polarization handling devices are needed. One example is for polarization diversity circuits like the ones used for optical interconnects, when edge coupling is employed. An important task of polarization diversity circuits is to handle polarization dependent loss and polarization mode dispersion which comes from the polarization dependence of the large birefringence in the silicon-on-insulator (SOI) platform. Another use of it is in separating the different polarizations, for example, in a coherent optical receiver. These functions are handled using devices like polarization beam splitters (PBS) and polarization rotators [1,2].

Another simpler approach is to design the circuit for one polarization [3] and avoid the other with a device like a TE- or TM-pass polarizer. In such a case, polarization division multiplexing cannot be exploited, but the circuit is simplified. It is well known that dispersion compensating filters [4] in integrated optics only work with one polarization, mainly TE. Normally, to introduce only one polarization into the circuit a bulk circulator and PBS are used [5–7]. When the TE mode is desired, then, a TE-pass polarizer can be placed at the input of the dispersion compensating filter avoiding both the PBS and the circulator. Consequently, a simple and compact TE-pass polarizer is needed.

Other important applications for pass polarizers are in the domain of quantum communications, optical sensing, and polarization filters. It is advantageous that the integrated TE-pass polarizer is Complementary metal–oxide–semiconductor (CMOS) compatible to leverage existing nanofabrication processes, mass production, and also be compatible with silicon photonic integrated circuits. With CMOS compatibility, we refer to the possibility of production in a state-of-the-art CMOS foundry [8]. The desired characteristics of a TE-pass polarizer are a high polarization extinction ratio (*PER*), low insertion loss (IL) and, depending on the application, a broad bandwidth. A typical value for the *PER* is roughly 20 dB, for the IL less than 2-3 dB and the bandwidth more than 150-200 nm [9]. The wide bandwidth is not only important for the coherent optical receiver or to handle WDM and OFDM signals, but also, it will be useful if the TE-pass polarizer can work in all the well-known bands exploited by silicon photonics (O, E, C, S, and L) with the same simple structure. There are many applications of silicon photonics in the O, E, C and L [10,11] bands. Other applications -includes the S and U bands [12] in wavelength converters.

Over the years, several CMOS compatible integrated polarizers have been proposed. They have either been all dielectric [2,13] or plasmonic. The advantage of plasmonic polarizers is that they are typically more compact. On the other hand, plasmonic-based integrated polarizers suffer from higher losses due to the absorption of the optical field by the metal. Furthermore, the most common materials for plasmonics are silver (Ag) or gold (Au) are not CMOS compatible [9]. Many TE- or TM-pass polarizers using those non-CMOS compatible materials have been previously studied [15–21].

There are many plasmonic polarizers [15–21] that are excited form a photonic waveguide. Nevertheless, it will be interesting that a plasmonic polarizer can also be excited from a plasmonic waveguide [22] using the same structure as used for the photonic circuit. A photonic circuit is a circuit in which the devices are connected by photonic waveguides, like the strip or slab silicon (Si) photonics waveguide. In the same way, a plasmonic circuit refers to a circuit in which the different devices in the circuit are connected by plasmonic waveguides, like the hybrid plasmonic waveguide (HPW).

Many CMOS compatible integrated plasmonic polarizers have been recently shown. The TE-pass polarizer in [3] is an experimental demonstration that has a high measured insertion loss of around 4 dB when excited from a standard Si waveguide. However, when excited from a horizontal nano-plasmonic slot waveguide the measured insertion loss is around 2.2 dB. One disadvantage of this structure is that it has a difficult fabrication since some fabrication features are below 193 nm deep-UV lithography [8] limit. In [23], a TM-pass polarizer compatible with standard CMOS fabrication process is demonstrated. Nevertheless, it has a moderate polarization extinction ratio of around 13 dB. Continuing with the recent trend of using new materials for plasmonics [14] the CMOS compatible TE-pass polarizer analyzed in [24] offers a polarization extinction ratio of 20 dB with a very low insertion loss of around 0.0635 dB in a very compact footprint with a length of 3.5 μm. Nevertheless, this device also has dimensions below the minimum resolution of 193 nm deep-UV lithography [8] and adds new materials in the fabrication. The device has a bandwidth of 150 nm for an extinction ratio better than 15 dB [24] extending the length to 5.5 μm. The work presented in [25] has a limited bandwidth. Finally, in [26] a bandwidth of 300 nm is reported, but the device is excited from a slot waveguide.

Most of them have in common that they only operate in one band or they have limited bandwidth. To the best of our knowledge there is no TE-pass polarizer capable of working across the different optical communications bands O (1.26 µm - 1.36 µm), E (1.36 µm - 1.46 µm), S (1.46 µm - 1.53 µm), C (1.53 µm - 1.565 µm), L (1.565 µm - 1.625 µm), and U (1.625 µm - 1.675 µm) with the same simple structure and being able to be excited from a photonic and/or a plasmonic waveguide. Most of the designs, except [3], are only excited from a photonic waveguide.

In this work, a TE-pass polarizer based on a vertical hybrid plasmonic waveguide (HPW) [27] is presented. The metal of the HPW is segmented to diffract the TM mode and pass the TE mode. Such a TE-pass polarizer is CMOS compatible and can operate in several bands of a photonic circuit (excited by a Si strip waveguide) with a polarization extinction ratio better than 20 dB and the insertion losses below 1.7 dB over all the optical communication bands. The same concept can also be excited from an HPW to be exploited in a plasmonic circuit.

## 2. Operational principle and design decisions

In this section, the operational principle of the device, the design decisions, and the optimization process are described. The first decision was to use an HPW [21] since it is one of the most promising candidates for a plasmonic circuit [28].

A TE-pass polarizer can be done with an HPW with a continuous metal. In this kind of waveguides, the TM fundamental mode, like the one in Fig. 1(a), has a higher absorption loss than the TE mode, like the one represented in Fig. 1(b) due to the metal.

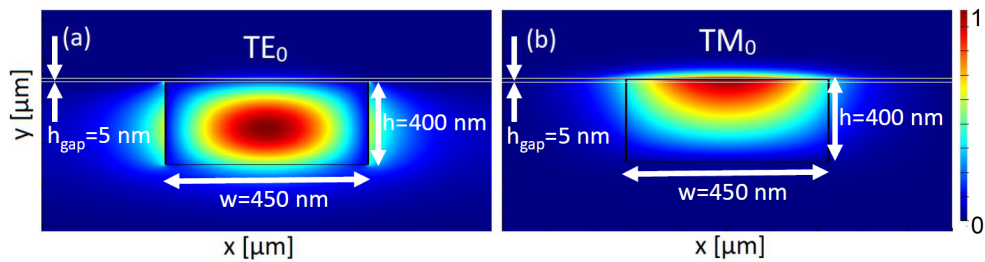

Fig. 1. Example of an electric field of the $TE_0$ (a) and $TM_0$ mode (b) in an HPW. The slot is made of $SiO_2$. The $TE_0$ mode is concentrated in the Si core while the $TM_0$ mode is concentrated in the slot and the interface between the metal and the slot. The wavelength is 1550 nm.

Hence, the $TM_0$ to extinguish more than the $TE_0$ mode as they propagate along the HPW. Nevertheless, due to the not so big difference in absorption losses between $TE_0$ and $TM_0$, this would result in a long TE-pass polarizer of several mm.

For reducing the length of such a TE-pass polarizer, in this work, the metal was segmented to diffract the $TM_0$ mode. The $TM_0$ mode is confined mostly close to the metal of the waveguide-like in Fig. 1(b) and is highly affected by the grating. On the other hand, $TE_0$ is far from the grating, and it is not highly affected.

If the metal is segmented, the TM mode diffracts due to the grating effect. On the other hand, the TE mode is relative away from the metal, and it is not affected by the grating.

The structure of the proposed device is represented in Fig. 2. The TE-pass polarizer to work in the plasmonic circuit represented in Fig. 2(a) is excited by an input HPW and the light is collected by an output HPW, between them, the metal was segmented to form the TE-pass polarizer structure. On the other hand, for the device to work in a photonic circuit, it is excited by an input Si strip waveguide and the light is collected by an output Si strip waveguide. Between them, the TE-pass polarizer consists of an HPW with the metal segmented.

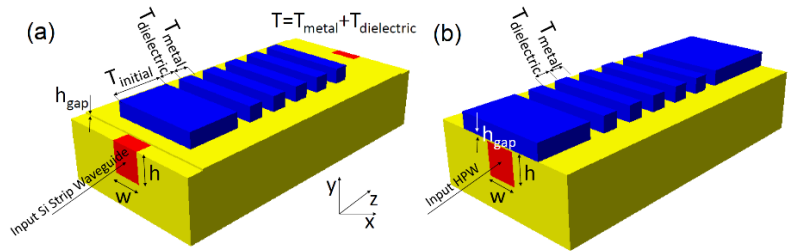

Fig. 2. Structure of the TE-pass polarizer. In (a) excited by a plasmonic waveguide. In (b) excited by a photonic waveguide.

This diffraction effect produced by the segmentation of the metal in the device is depicted in Figs. 3(a) and 3(b) for a plasmonic circuit and in Figs. 3(c) and 3(d) for a photonic circuit. These Figs. represent the complete optimized TE-pass polarizer whose optimization will be detailed in the next section.

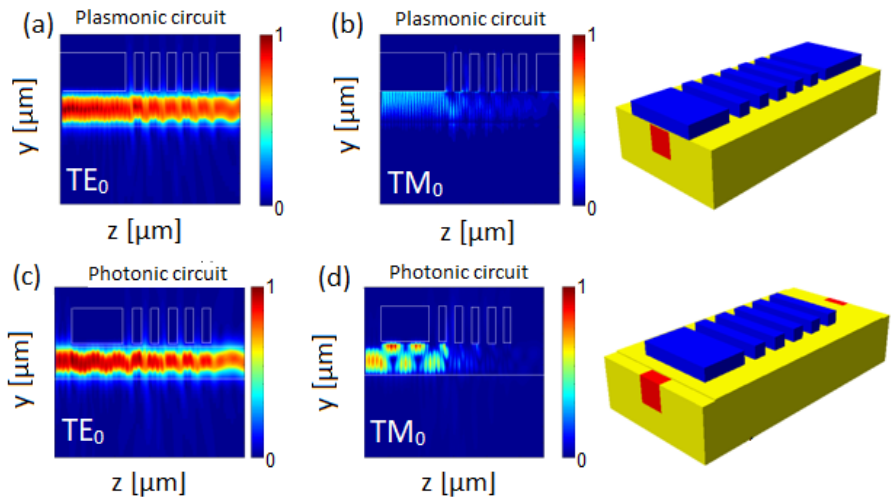

Fig. 3. The electric field of the $TE_0$ mode passing through the TE-pass polarizer (a) in a plasmonic circuit. In (b), the electric field of the $TM_0$ mode being filtered by the TE-pass polarizer. In (c), there is the electric field of the $TE_0$ mode passing the TE-pass polarizer while in (d) the $TM_0$ mode is filtered. The wavelength is 1550 nm.

In Fig. 3, the $TE_0$ mode passes while the $TM_0$ mode does not. The parameters of the TE-pass polarizer in Fig. 3 were optimized as will be described in the next section.

In the following section, it is explained how the optimization of the TE-pass polarizer was done regarding materials and parameters.

## 3. Optimization of the TE-pass polarizer

In this section, we present the optimization process done to design the structure. As a first step, we selected the material for the structure, then, we optimized the parameters of the structure ($w$, $h$, $h_{gap}$ and $T_{metal}$). To simplify we selected $T_{metal} = T_{dielectric}$.

Regarding the selection of the materials, as the core of the waveguide, we selected Si to be compatible with the Silicon Photonics platform [8]. To be CMOS compatible, and leverage existing microelectronic fabrication processes, we need to select between Aluminum (Al) and Copper (Cu) for the metal [8] of the HPW. Since Cu has less optical losses than Al [10,11], we selected Cu. Regarding the diffusion barrier for the slot, there are many possibilities: Ta, TaN, Ti, TiN, $Si_3N_4$, and $SiO_2$ [29].

To select the material, we performed 2D simulations with MODE solutions of Lumerical [30]. The analysis was performed in the cross section of an HPW to try to maximize the difference in optical losses $\Delta\alpha_{eff}$ between the absorption loss of the $TM_0$ mode $\alpha_{eff,TM}$ and the absorption loss of the $TE_0$ mode $\alpha_{eff,TE}$. The value $\Delta\alpha_{eff} = \alpha_{eff,TM} - \alpha_{eff,TE}$ is related to the polarization extinction ratio of the device (PER) by multiplying $\Delta\alpha_{eff}$ by the length of the TE-pass polarizer $L$ (when the metal is continuous). The polarization extinction ratio is defined as the ratio of the power in the $TE_0$ mode at the output of the TE-pass polarizer $P_{out,TE}$ divided by the power of the $TM_0$ mode at the output $P_{out,TM}$ of such a polarizer. The final expressions is PER = $P_{out,TE}/P_{out,TM}$. The final relation is PER = $\Delta\alpha_{eff} \cdot L$. Hence, a large $\Delta\alpha_{eff}$ is desired.

On the other hand, the parameter $\alpha_{eff,TE}$ is related to the insertion loss (IL) of the device when it is multiplying by the length of the polarizer $L$ (when the metal is continuous). The insertion loss IL = $P_{out,TE}/P_{in,TE}$ is defined as the power in the TE mode at the output $P_{out,TE}$ divided by the power of the TE mode at the input $P_{in,TE}$. The final relation is IL = $\alpha_{eff,TE} \cdot L$. Hence, a small value of $\alpha_{eff,TE}$ is desired.

Consequently, for the optimization of the cross-section of the HPW, we selected the Figure of Merit (FoM) equal to FoM = $\Delta\alpha_{eff}/\alpha_{eff,TE}$ to maximize the polarization extinction ratio and reduce the insertion loss of the TE-pass polarizer.

The main parameters that influence $\alpha_{eff,TM}$ and $\alpha_{eff,TE}$ are the width of the Si core $w$, the height of the Si core $h$ and the thickness of the slot $h_{gap}$ as shown in Fig. 2. Consequently, we studied the FoM = $\Delta\alpha_{eff}/\alpha_{eff,TE}$ versus the parameters $w$, $h$, and $h_{gap}$. For fabrication reasons, we compared the slots with the following materials: $SiO_2$, $Si_3N_4$, Ta, and Ti [29]. The analysis is presented in Fig. 4.

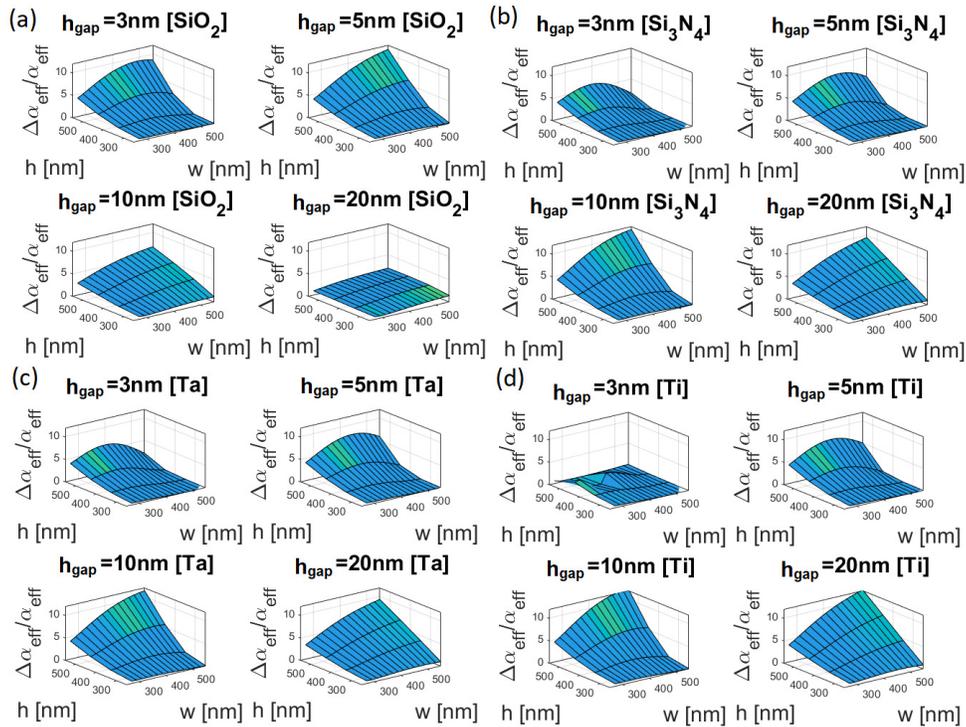

Fig. 4. Study of the FoM = $\Delta\alpha_{eff}/\alpha_{eff,TE}$ for the different parameters of the HPW $w$, $h$ and $h_{gap}$ with a $SiO_2$ (a), $Si_3N_4$ (b), Ta (c) and Ti (d). Typical parameters for $w$, $h$ and $h_{gap}$ are used.

In Figs. 4(a)–4(d) we scanned the parameter $h_{gap}$ from 3 nm (minimum fabrication thickness [10,11,29]) to 20 nm until a maximum is reached for an intermediate $h_{gap}$. The parameters *w* and *h* are scanned from 250 nm to 500 nm which are common values for the HPW [10,29].

In Fig. 4(a) the FoM = $\Delta\alpha_{eff}/\alpha_{eff,TE}$ is represented versus *w* and *h* for four different $h_{gap}$ values: 3 nm, 5 nm, 10 nm and 20 nm. It is possible to observe that the maximum FoM is reached for $h_{gap}$ = 5 nm. The maximum value of the FoM is 10.

In the case of Fig. 4(b) with the slot of $Si_3N_4$, the same study is done, and a maximum value of the FoM = 10 is reached for $h_{gap}$ = 10 nm. Finally, for Ta in Fig. 4(c) and Ti in Fig. 4(d) the maximum value reached is for $h_{gap}$ = 10 nm in both cases.

In all cases, in Fig. 4 the FoM increases when *w* and *h* are increased. Since the $TE_0$ mode is concentrated in the Si core, if we increase *h*, the mode is farther away from the metal which will reduce the optical losses of the $TE_0$ mode $\alpha_{eff,TE}$. This will increase the FoM = $\Delta\alpha_{eff}/\alpha_{eff,TE}$. On the other hand, in a waveguide when you increase *w* the modes have less optical losses, increasing *w* reduces $\alpha_{eff,TE}$ so that $\Delta\alpha_{eff}/\alpha_{eff,TE}$ increases with *w*. On the other hand, the $TM_0$ mode loss strongly depends on $h_{gap}$ [10]. On the contrary, the losses of the $TE_0$ mode does not depend strongly on $h_{gap}$ because the field is not confined in the slot. The relation between the optical loss of the mode and the effective index of the material used can be found in [31].

Finally, for fabrication reasons, we have decided to employ $SiO_2$ in the rest of the study. Although using $Si_3N_4$, Ta, or Ti may result in a similar performance. Selecting $SiO_2$ in the slot will fix the parameter to $h_{gap}$ = 5 nm to obtain the maximum FoM. Now, it will be necessary to fix *w* and *h*. If we select the biggest values for *w* and *h* that are 500 nm, then the waveguide will be highly multimode. To avoid that, we fixed *w* = 450 nm and *h* = 400 nm. With such dimensions, there is only $TE_0$ and multimode TM modes ($TM_0$, $TM_1$, and $TM_2$). High order TM modes will also be filtered by the structure. The final optimized parameters of the cross section are *w* = 450 nm, *h* = 400 nm, and $h_{gap}$ = 5 nm ($SiO_2$). The optimization of $T_{metal}$ is the last parameter to be optimized.

## 4. Optimization of $T_{metal}$ in the TE-pass polarizer

In this section, we study the performance of the TE-pass polarizer with the HPW previously optimized and with continues metal. Later, we segmented the metal of the HPW, and the parameter $T_{metal}$ is optimized.

For the optimized values of the cross-section of the HPW the effective index of the photonic TE fundamental mode is $n_{eff,TE}$ = 2.75, and the absorption loss is $\alpha_{eff,TE}$ = 0.0411 dB/μm. Regarding the plasmonic TM fundamental mode the effective index is $n_{eff,TM}$ = 3.21 and the absorption loss is $\alpha_{eff,TM}$ = 0.2370 dB/μm. To calculate the polarization extinction ratio, we use the formula PER = $(\alpha_{eff,TM} - \alpha_{eff,TE}) \cdot L$ and for the insertion loss, we use the formula IL = = $(\alpha_{eff,TE}) \cdot L$, where *L* is the length of the HPW. The results are shown in Fig. 5.

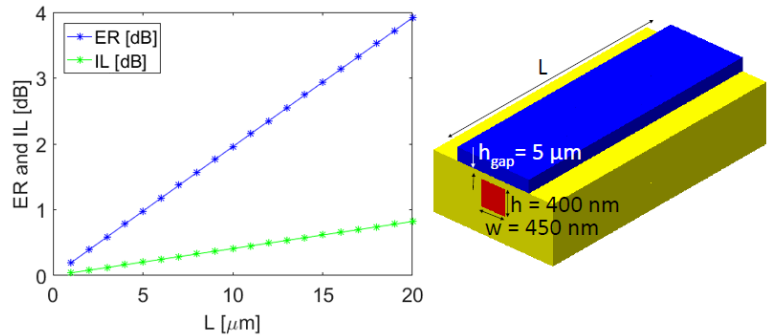

Fig. 5. Estimated polarization extinction ratio and insertion loss of a TE-pass polarizer with continues metal which consists of an HPW of length L. The dimensions of the HPW are *w* = 450 nm, *h* = 400 nm and $h_{gap}$ = 5 nm ($SiO_2$). The wavelength is 1550 nm.

With this device, 20 μm are needed to obtain an extinction ratio of 4 dB. By segmenting the metal, the device can be shortened.

To optimize the last parameter $T_{metal}$, it is needed to calculate the length of the TE-pass polarizer and the optimum size of the metal segments $T_{metal}$. To do this, we performed 3D FDTD simulations using Lumerical [30].

For the initial simulations, we selected minimum value for $T_{metal}$ around 0.3 μm to use lift-off in the fabrication process. On the other hand, the maximum value of $T_{metal}$ was set to 1 μm to have a compact TE-pass polarizer still. To further complete the scanning, we also selected an intermediate value of $T_{metal}$ around 0.5 μm. To agree with the grating theory, we selected the period of the grating $T = T_{metal} + T_{dielectric}$, with $T_{metal} = T_{dielectric}$.

Using the $FoM = \Delta\alpha_{eff} \cdot L / \alpha_{eff,TE} \cdot L = PER/IL$ as previously explained, the result of the 3D FDTD simulation is presented in Fig. 6(a).

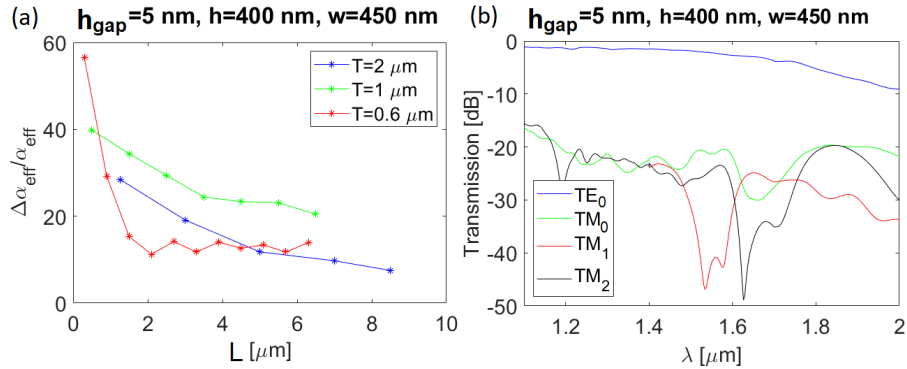

Fig. 6. Estimation of the $FoM = PER/IL$ of the TE-pass polarizer with a segmented metal of $T_{metal}$ (a) The wavelength is fixed to 1550 nm and the HPW have the optimized values. In (b) the spectrum of the TE-pass polarizer with $L = 5.5$ μm and $T_{metal} = 0.5$ μm.

In Fig. 6(a), the maximum of the $FoM$ is for $T = 0.5$ μm. As one of the requirements for a TE-pass polarizer is to have a polarization extinction ratio around 20 dB, we selected a minimum length of $L = 5.5$ μm. The final parameters are summarized in Table 1.

Table 1. Summary of the optimized parameters of the TE-pass polarizer with the HPW.

| Parameter: | $w$ | $h$ | $h_{gap}$ | $T_{metal}$ | $L$ |
|---|---|---|---|---|---|
| Value: | 450 nm | 400 nm | 5 nm | 0.5 μm | 5.5 μm |

With the values summarized in Table 1, we calculated the spectrum of the TE-pass polarizer with the segmented metal for the wavelength window from 1.1 μm to 2 μm. The estimation is shown in Fig. 6(b). In such a case, we included the higher order plasmonic TM modes named $TM_1$ and $TM_2$. It is possible to observe that $TM_1$ and $TM_2$ are also filtered.

To the propagation of $TM_1$ and $TM_2$ in represented it in Figs. 7(a) and 7(b).

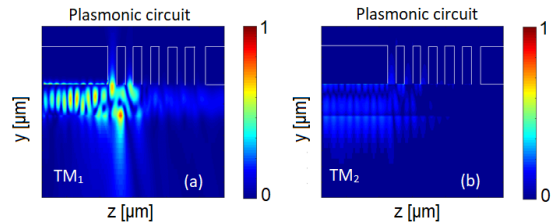

Fig. 7. Filtering of $TM_1$ (a) and $TM_2$ (b) for the optimized TE-pass polarizer with $w = 450$ nm, $h = 400$ nm, $h_{gap} = 5$ nm ($SiO_2$), $T_{metal} = 0.5$ μm, and $L = 5.5$ μm for 1550 nm.

The optimized performance of the TE-pass polarizer is summarized in Table 2.

Table 2. Performance of the TE-pass polarizer in a plasmonic circuit.

| Band: | O | E | S | C | L | U |
|---|---|---|---|---|---|---|
| Polarization extinction ratio: | 20 dB | 20 dB | 21 dB | 20 dB | 18 dB | 22 dB |
| Insertion loss: | 0.7 dB | 1 dB | 1.3 dB | 1.3 dB | 1.5 dB | 1.7 dB |
| Wavelength: | 1.31 μm | 1.4 μm | 1.5 μm | 1.55 μm | 1.6 μm | 1.65 μm |

In the C-band, the polarization extinction ratio is around 20 dB for an IL = 1.3 dB at 1.55 μm for 0.5 μm which is better than using a continuous metal.

As a summary, we optimized the cross-section of the HPW of the TE-pass polarizer using 2D simulations solutions [30]. With this, we calculated $w$ = 450 nm, $h$ = 400 nm and $h_{gap}$ = 5 nm (SiO$_2$). After that, we performed 3D FDTD simulations to calculate the parameter $T_{metal}$ = 0.5 μm and $L$ = 5.5 μm to obtain a polarization extinction ratio better than 20 dB at 1550 nm.

It will be interesting to check the simulations by scanning the cross-section of the HPW using 3D FDTD simulations to see if they agree with the 2D simulations. For this, we fixed $h_{gap}$ = 5 nm (SiO$_2$), $T_{metal}$ = 0.5 μm, and $L$ = 5.5 μm. And we scanned $w$ and $h$. The results of the 3D FDTD estimation is represented in Fig. 8. To scan all the parameters in 3D is not computationally affordable in a reasonable time.

In Fig. 8, the trend is the same as in Fig. 4(a). Consequently, we have validated the 2D approximation we did. It means when we increase $w$, $FoM = \Delta\alpha_{eff}/\alpha_{eff,TE}$ increases. It has the same trend as Fig. 4(a). In the same manner, when we increase $h$, $FoM = \Delta\alpha_{eff}/\alpha_{eff,TE}$ increases.

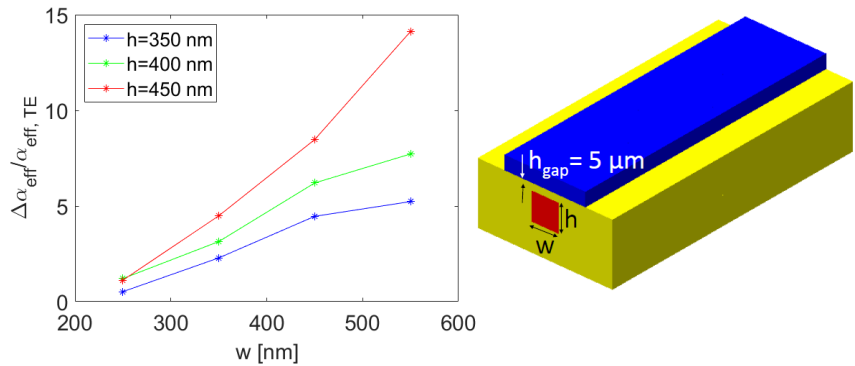

Fig. 8. Scanning $w$ and $h$ in 3D FDTD for $h_{gap}$ = 5 nm (SiO$_2$), $T_{metal}$ = 0.5 μm, and $L$ = 5.5 μm at 1550 nm. This scan is performed to know if the results are like the results in 2D represented in Fig. 4(a).

To conclude this section, we have analyzed the performance of the TE-pass polarizer with segmented metal for different optical bands in a plasmonic circuit. Now, it will be interesting to exploit the same device in a photonic circuit from which the TE-pass polarizer will be excited from a photonic strip waveguide.

## 5. Performance of TE-pass polarizer in a photonic circuit

In this section, we calculate the optimization of the TE-pass polarizer to be excited from a photonic waveguide. The structure is presented in Fig. 1(b) and it consists of an input Si strip waveguide to excite the device, the TE-pass polarizer consisting of segmented metals and an output Si strip waveguide to collect light. The dimensions of the Si strip waveguide employed are $w$ = 450 nm and $h$ = 400 nm. The modes of such waveguide are represented in Fig. 9.

The TE$_0$ mode is represented in Fig. 9(a) and it will be passing the TE-pass polarizer. On the other hand, the TM modes in Figs. 4(b)–4(d) will be filtered by the device.

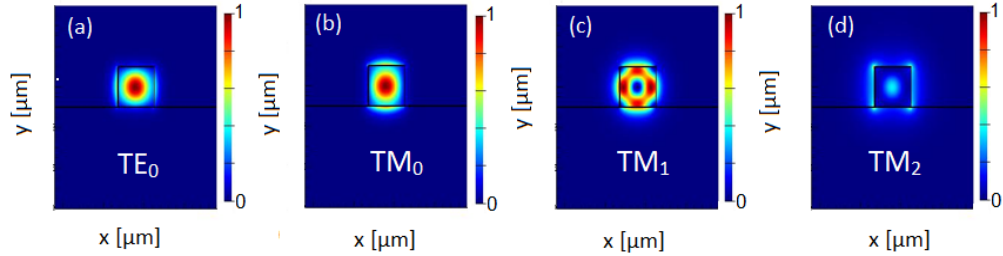

Fig. 9. Modes supported by a Si strip waveguide of $w$ = 450 nm and $h$ = 400 nm. In (a) the electric field of $TE_0$, in (b) the $TM_0$ mode. $TM_1$ in (c) and $TM_2$ in (d).

In this section, we employed the same optimized values for the device. It means, the parameters of the cross-section of the HPW are $w$ = 450 nm, $h$ = 400 nm, and $h_{gap}$ = 5 nm ($SiO_2$). The metal was segmented with $T_{metal}$ = 0.5 μm. Regarding the dimension of the input/output strip waveguides, we have selected $w$ = 450 nm and $h$ = 400 nm to butt-couple it to the device.

The bandwidth of the TE-pass polarizer is represented in Fig. 10(a). For comparison, in Fig. 10(b) we placed the bandwidth of a TE-pass with continuous metal using the same dimensions.

By comparing Figs. 10(a) and 10(b) we can observe that the polarization extinction ratio of the TE-pass polarizer with segmented metal is larger for a similar insertion loss at 1550nm. Consequently, by segmenting the metal, the polarization extinction ratio is incremented while reducing the length of the device.

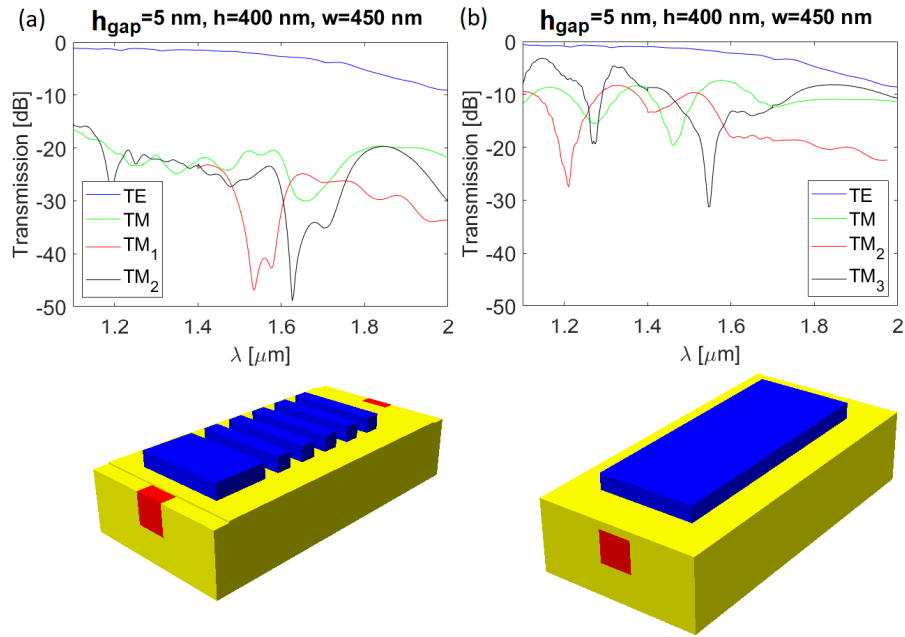

Fig. 10. Bandwidth of the TE-pass polarizer with segmented metal with the optimized values (a) and the same TE-pass polarizer with an HPW of continuous metal (b).

As shown before, the $TE_0$ mode of the Si strip waveguide passes the structure as represented in Fig. 3(c) and the fundamental TM mode of the Si strip waveguide is filtered by the device as shown in Fig. 3(d). On the other hand, the higher order TM modes, $TM_1$ and TM$_2$, supported by the input Si strip waveguide are also filtered. This is exemplified in Fig. 11.

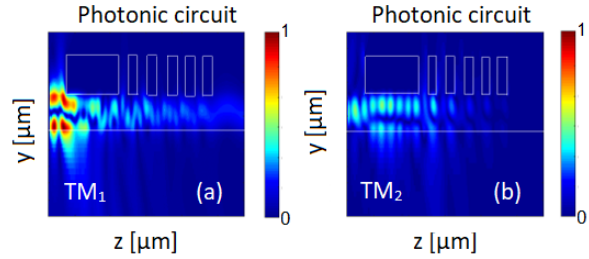

Fig. 11. Optimized TE-pass polarizer with segmented metal for a photonic circuit for TM$_1$ (a) and TM$_2$ (b) at 1550 nm.

The performance is summarized in Table 3 for the different optical communication bands.

Table 3. Performance of the TE-pass polarizer in the different optical bands.

| Band | O | E | S | C | L | U |
|---|---|---|---|---|---|---|
| Polarization extinction ratio | 20.5 dB | 21 dB | 20.9 dB | 19.3 dB | 19.5 dB | 27.2 dB |
| Insertion loss | 1.5 dB | 1.5 dB | 1.6 dB | 2.2 dB | 2.5 dB | 2.8 dB |
| Wavelength | 1.31 μm | 1.4 μm | 1.5 μm | 1.55 μm | 1.6 μm | 1.65 μm |

As a final step, we did a further analysis of the period in Fig. 12. We scanned the period $T$ in intermediate points to know if there is another better period. In Fig. 12(a) we presented the transmission of different periods between 0.5 μm and 1.5 μm. To compare all of them we calculated the *FoM* in Fig. 12(b). It is possible to observe that $T = 1$ μm has the best behavior regarding the bandwidth. On the other hand, it is only better by $T = 0.5$ μm in the C-band. Nevertheless, for $T = 0.5$ μm we have a $T_{metal} = 0.25$ μm which is at the limit of the lift-off resolution.

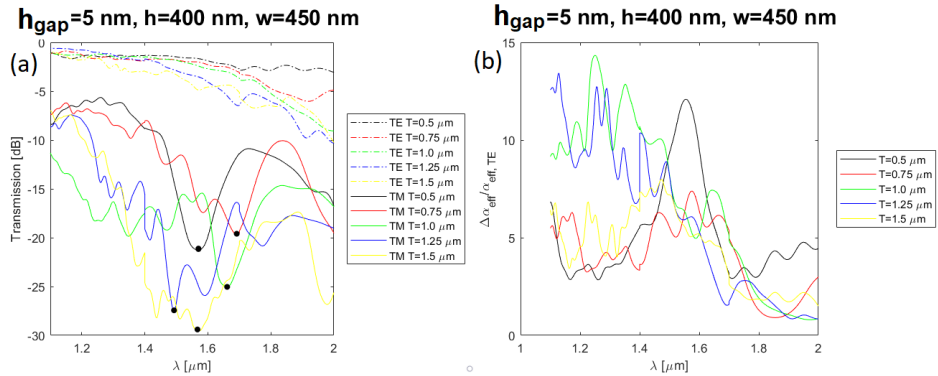

Fig. 12. Scanning the period $T$ of the optimized plasmonic TE-pass polarizer. In (a) the transmission of the TE and TM modes and in (b) the FoM = $\Delta\alpha_{eff}/\alpha_{eff,TE}$. The dots in Fig. 12(a) are the points selected to study the grating behavior.

From Fig. 12(a) it is possible to observe that the TM mode has a valley that depends on the wavelength. We marked it with black dots. We checked that the valley shifts according to the following formula:

$$T_{estimated} = \frac{m \cdot \lambda_{valley}}{2 \cdot n_{eff}(\lambda_{valley})} \quad (1)$$

Where $T_{estimated}$ is the estimated period of the grating, m is the order of the grating, $\lambda_{valley}$ is the wavelength of the valley in transmission, and $n_{eff}$ is the effective index of $TM_0$ mode at the wavelength $\lambda_{valley}$. The results are summarized in Table 4. It is possible to observe that the simulations follow the grating behavior. This confirms that the device is diffracting the $TM_0$ mode. The same happens with $TM_1$ and $TM_2$.

Table 4. Estimation of the period *T* of the grating.

| T [μm] | $\lambda_{valley}$ [μm] | $n_{eff}$ | m | $T_{estimated}$ [μm] |
|---|---|---|---|---|
| 0.5 | 1.575 | 2.91 | 2 | 0.54 |
| 0.75 | 1.63 | 2.83 | 3 | 0.84 |
| 1.0 | 1.55 | 2.85 | 4 | 1.08 |
| 1.25 | 1.475 | 2.99 | 5 | 1.23 |
| 1.5 | 1.56 | 2.91 | 6 | 1.59 |

## 6. Comparison with the state-of-the-art

In this section, the proposed device is compared with the CMOS compatible plasmonic TE-pass polarizers reported until now. The main parameters, like the polarization extinction ratio, the insertion loss, and the bandwidth are discussed. The parameters are summarized in Table 5.

In conclusion, our structure offers a larger extinction ratio and a smaller insertion loss than the experimental work presented in [3].

Table 5. Comparison with the state-of-the-art of CMOS compatible TE-pass polarizers at 1.55μm.

| Group: | A*STAR [3] (2013) | Mansoura [24] (2016) | Southeast [25] (2016) | Huazhong [26] (2016) | Us |
|---|---|---|---|---|---|
| Polarization extinction ratio | 16 dB | 20 dB | 34.74 dB | 41.62 dB | 19.3 dB |
| Insertion loss | 4 dB* | 0.06350 dB | 0.43 dB | 0.38 dB | 2.2 dB |
| Length | 4 μm | 3.5 μm | 8 μm | 3 μm | 5.5 μm |
| Bandwidth | 80 nm | 150 nm** | 250 nm | 300 nm | 500 nm (E, S, C, L, S and U bands***) |
| Type of work | Experimental | Theoretical | Theoretical | Theoretical | Theoretical |

*From [1] when excited from a Si strip waveguide.

** From Fig. 4(a) [24] when the polarization extinction ratio is 15 dB (5 dB less than the maximum of 20 dB). The device is 3.5 μm, but we took 5.5 μm (like ours), and we scaled the ER and IL linearly.

***For a polarization extinction ratio bigger than 22 dB.

On the other hand, the work presented by Mansoura [24] has an excellent trade-off regarding the polarization extinction ratio and the insertion loss. Nevertheless, it has a limited bandwidth of 150 nm (for a polarization extinction ratio bigger than 15 dB and a length of 5.5 μm). Furthermore, this structure presents features smaller than the minimum resolution of 193 nm deep-UV lithography.

The work in [25] has limited bandwidth, and the structure is not simple to fabricate since it involves the fabrication of a thin layer of ITO in the middle of the strip Si waveguide. Finally, the device proposed in [26] has an excellent trade-off between *PER* and *IL* but the bandwidth is limited, and it is excited from a slot waveguide.

One of the advantages of our device is that it offers the possibility of working in the O, E, S, C, L, and U bands with an acceptable polarization extinction ratio and moderate insertion loss with the same simple structure. Such a concept can be exploited both in a plasmonic circuit and in a photonic circuit. This concept can be extended to TM-pass polarizers like in [32].

## 7. Conclusion

In this work, the design and optimization of a novel plasmonic CMOS compatible TE-pass polarizer were presented. To design the device, we optimized its most important parameters using 2D and 3D FDTD simulations. The device works in several optical bands with the same simple CMOS compatible structure being the main novelty of the device the segmented metal of the HPW. This allows to pass of the TE mode and blocks the TM modes. This TE-pass polarizer can work in the traditional communication bands for silicon photonics: O, E, S, C, L, and U with the same structure.

## Funding

EPSRC (EP/P006973/1).

## Acknowledgments

We would like to acknowledge CMC Microsystems for the provision of products and services that facilitated this research, including Lumerical. We are grateful for support from the Future Compound Semiconductor Manufacturing Hub (CS Hub) funded by EPSRC grant reference EP/P006973/1.